\documentclass[aps,pra,twocolumn,superscriptaddress,floatfix]{revtex4}
\usepackage{amsmath,amssymb,graphicx,graphics,color}
\usepackage{dcolumn}
\usepackage{bm}
\usepackage{psfrag}

\newcommand{\bs}{\boldsymbol}

\begin{document}

\title{Ultra-cold bosons in zig-zag optical lattices}
\author{S. Greschner}
\affiliation{Institut f\"ur Theoretische Physik, Leibniz Universit\"at Hannover, 30167~Hannover, Germany}
\author{L. Santos}
\affiliation{Institut f\"ur Theoretische Physik, Leibniz Universit\"at Hannover, 30167~Hannover, Germany}
\author {T. Vekua}
\affiliation{Institut f\"ur Theoretische Physik, Leibniz Universit\"at Hannover, 30167~Hannover, Germany}

\begin{abstract}
Ultra-cold bosons in zig-zag optical lattices present a rich physics 
due to the interplay between frustration, induced by lattice geometry, 
two-body interaction and three-body constraint. Unconstrained bosons 
may develop chiral superfluidity and a Mott-insulator even at vanishingly 
small interactions. Bosons with a three-body constraint allow 
for a Haldane-insulator phase in non-polar gases, as well as 
pair-superfluidity and density wave phases for attractive interactions.
These phases may be created and detected within the current state 
of the art techniques.
\end{abstract}

\maketitle

\date{\today}




\section{Introduction}

 Atoms in optical lattices offer extraordinary possibilities for the controlled emulation and analysis of lattice models and 
quantum magnetism~\cite{Lewenstein2007,Yukalov2009}. Various lattice geometries are attainable by means of proper laser arrangements, 
including triangular~\cite{Becker2010} and Kagome~\cite{Jo2012} lattices, 
opening fascinating possibilities for the study of geometric frustration, which may  
result in flat bands with the constrained mobility and largely enhanced role of interactions~\cite{Huber2010}. 
Moreover, the value and sign of inter-site hopping may be modified by means of shaking techniques~\cite{Eckardt2005,Zenesini2009}, 
allowing for the study of frustrated antiferromagnets with bosonic lattice gases~\cite{Struck2011}.

Interatomic interactions may be controlled basically at will by means of Feshbach 
resonances~\cite{Chin2010}. In particular, large on-site repulsion may allow for the suppression of double 
occupancy in bosonic gases at low fillings~(hard-core regime). Interestingly, it has been recently suggested that, due to a 
Zeno-like effect, large three-body loss rates may result in an effective three-body constraint, in which no more than 
two bosons may occupy a given lattice site~\cite{Daley2009}. This constraint opens exciting novel 
scenarios, especially in what concerns stable Bose gases with attractive on-site interactions, 
including color superfluids in spinor Bose gases~\cite{Titvinidze2011} and pair-superfluid 
phases~\cite{Daley2009,Bonnes2011, Chen2011}. The suppression of three-body occupation has been hinted 
in recent experiments~\cite{Mark2012}.

Under proper conditions, lattice gases may resemble to a large extent effective spin models, e.g. hard-core bosons may 
be mapped into a spin-$1/2$ XY Heisenberg model~\cite{Lewenstein2007}. 
Lattice bosons at unit filling resemble to a large extent spin-$1$ chains~\cite{DallaTorre2006}, and in the presence of inter-site interactions, as it is the case of polar gases~\cite{Lahaye2009}, have been shown 
to present a gapped Haldane-like phase~\cite{Haldane1983}~(dubbed Haldane-insulator~(HI)~\cite{DallaTorre2006,Berg2008}) 
characterized by a non-local string-order~\cite{DenNijs1989}.

In this work  we analyze the physics of ultra-cold bosons in zig-zag optical lattices. We show that the interplay 
of frustration and interactions lead to a different physics for unconstrained and constrained~(with up to two particles per site) 
bosons. For unconstrained bosons, geometric frustration induces chiral superfluidity, and allows for a Mott-insulator 
phase even at vanishingly small interactions. For constrained bosons, we show that a Haldane-insulator phase becomes possible 
even for non-polar gases. Moreover, pair-superfluid~\cite{Daley2009,Bonnes2011, Chen2011} and density-wave phases 
may occur for attractive on-site interactions. A direct first-order phase transition from Haldane-insulator to pair-superfluid 
is observed and explained. These phases may be realized and detected with existing state of the art techniques.

The structure of the paper is as follows. In Sec.~\ref{sec:Model} we 
introduce the zig-zag lattice model under consideration. Sec.~\ref{sec:Unconstrained} is devoted to the unconstrained case, whereas Sec.~\ref{sec:Constrained} deals with bosons with a two-body constraint.
Finally, in Sec.~\ref{sec:conclusions} we summarize our conclusions. Further details on analytical and numerical procedures are discussed in the Appendices.

\begin{figure}[t]
\includegraphics[width=5.0cm]{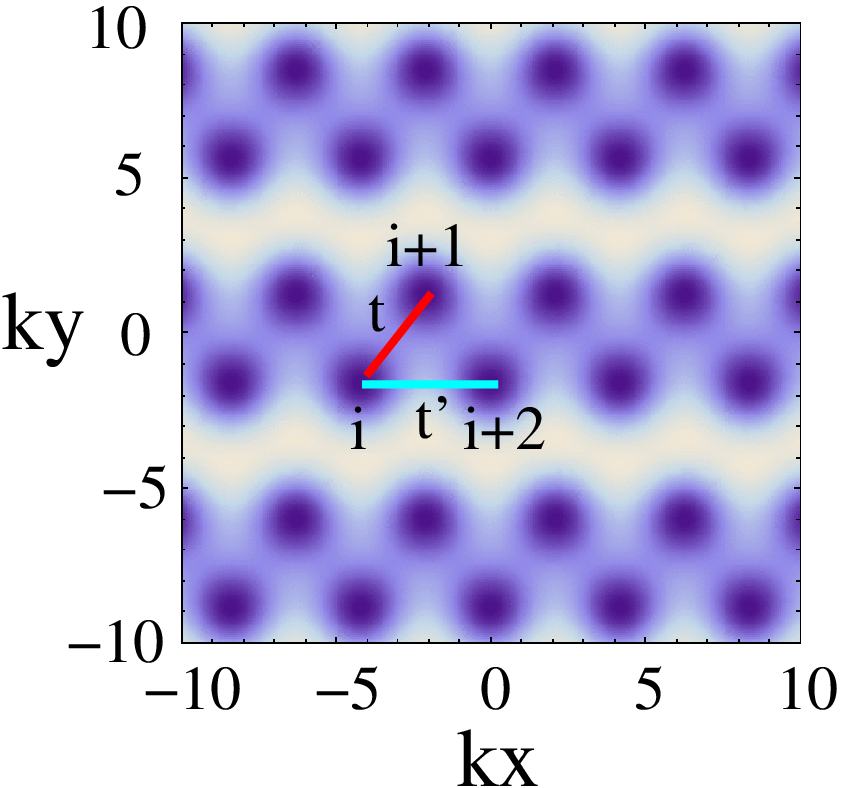}
\caption{Zig-zag chains formed by an incoherent superposition between a triangular lattice~\cite{Becker2010}
$V_1({\vec r}\equiv(x,y))=V_{10}\left [ \sin^2\left ({\vec b}_1\cdot{\vec r}/2\right )+\sin^2\left ({\vec b}_2\cdot {\vec r}/2\right )
+\sin^2\left (({\vec b}_1-{\vec b}_2)\cdot{\vec r}/2\right )\right ]$, 
with $k$ the laser wavenumber, ${\vec b}_1=\sqrt{3}k {\vec e}_y$ and ${\vec b}_2=\sqrt{3}k(\sqrt{3}{\vec e}_x/2-{\vec e}_y/2)$, and an additional lattice $V_2({\vec r})=V_{20}\sin^2(\sqrt{3}ky/4 - \pi/4)$. In the figure, in which $V_{20}/V_{10}=2$, 
darker regions mean lower potential. The hopping rate between nearest~(next-nearest) neighbors is $t>$~($t'$).}
\vspace*{-0.3cm}
\label{fig:1}
\end{figure}



\section{Zig-zag lattices} 
\label{sec:Model}
\label{sec:intro}
 In the following we consider bosons in zig-zag optical lattices. As shown in Fig.~\ref{fig:1}, this particular geometry may result from the incoherent superposition of a triangular lattice with elementary cell vectors $\vec{a}_1=a {\vec e}_x$ and 
$\vec{a}_2=a(\frac{1}{2}{\vec e_x}+\frac{\sqrt{3}}{2}{\vec e}_y)$ (formed by three laser beams of wavenumber $k=4\pi/3a$  
oriented at $120$ degrees from each other, as discussed in Ref.~\cite{Becker2010}) and a superlattice 
with lattice spacing $\sqrt{3}a$ oriented along $y$. For a sufficiently strong superlattice, zig-zag ladders are formed, and the hopping between ladders may be neglected. We will hence concentrate in the following on the physics of bosons in a single zig-zag ladder, which is to a large extent given by the rates $t$ and $t'$ characterizing the hopping along the two directions ${\vec a}_{1,2}$~(Fig.~\ref{fig:1}). As shown in Ref.~\cite{Struck2011}, a periodic lattice shaking may be employed to control the 
value of $t$ and $t'$ independently.  Interestingly, their sign may be controlled as well. In the following we consider an inverted sign for both hoppings, which result in an anti-ferromagnetic coupling between sites~\cite{Struck2011}.

Ordering the sites as indicated in Fig.~\ref{fig:1}, the physics of 
the system is given by a Bose-Hubbard Hamiltonian with on-site 
interactions characterized by the coupling constant $U$, 
nearest-neighbor hopping $t<0$ and next-nearest-neighbor hopping $t'<0$:
\begin{eqnarray}
\label{eq:H-BH}
H&=&\sum_i \left [ -\frac{t}{2} b_i^\dag b_{i+1} - \frac{t'}{2} b_i^\dag b_{i+2} + {\rm H.c} \right ]\\
&+&\frac{U}{2}\sum_i n_i(n_i-1)+U_3\sum_i n_i(n_i-1)(n_i-2),\nonumber
\end{eqnarray}
where $b_i^\dag,b_i$ are the bosonic creation/annihilation operators 
of particles at site $i$, $n_i=b_i^\dag b_i$, and we have added the 
possibility of three-body interactions, characterized by 
the coupling constant $U_3$. We assume below an average unit filling 
${\bar n}=1$. 

\section{Unconstrained bosons} 
\label{sec:Unconstrained}
We discuss first the ground-state properties of unconstrained 
bosons~($U_3=0$). At $U=0$, 
the Hamiltonian~\eqref{eq:H-BH} is diagonalized in quasi-momentum 
space $H=\sum_k  \epsilon(k) b_k^\dag b_k$, 
with the dispersion $\epsilon(k)= |t| (\cos k +j\cos 2k)$, with $j\equiv t'/t$. 
Depending on the frustration $j$ we may distinguish two regimes.
If $j<1/4$, the dispersion $\epsilon(k)$ 
presents a single minimum at $k=\pi$, and hence small $U$ 
will introduce a superfluid~(SF) phase, with quasi-condensate at $k=\pi$.
If $j>1/4$, $\epsilon(k)$ presents two 
non-equivalent minima at $k=k0\equiv\pm \arccos[-1/4j]$. 
As shown below, interactions favor the predominant population of one of 
these minima, and the system enters a chiral superfluid~(CSF) phase 
with a non-zero local boson current characterized by a finite 
chirality~$\langle \kappa_i\rangle$, with 
$\kappa_i=\frac{i}{2}(b_{i}^\dag b_{i+1}-{\rm H.c.})$. At $j=1/4$, the 
Lifshitz point, the dispersion becomes quartic at the $k=\pi$ minimum, 
$\epsilon(k)\sim (k-\pi)^4$, the effective mass
$m=(\partial^2 \epsilon(k)/\partial k^2)_{k=\pi}^{-1}=1/t(1-4j)$ diverges, and 
even vanishingly small interactions become relevant.

\begin{figure}[t]
\includegraphics[width=8.5cm]{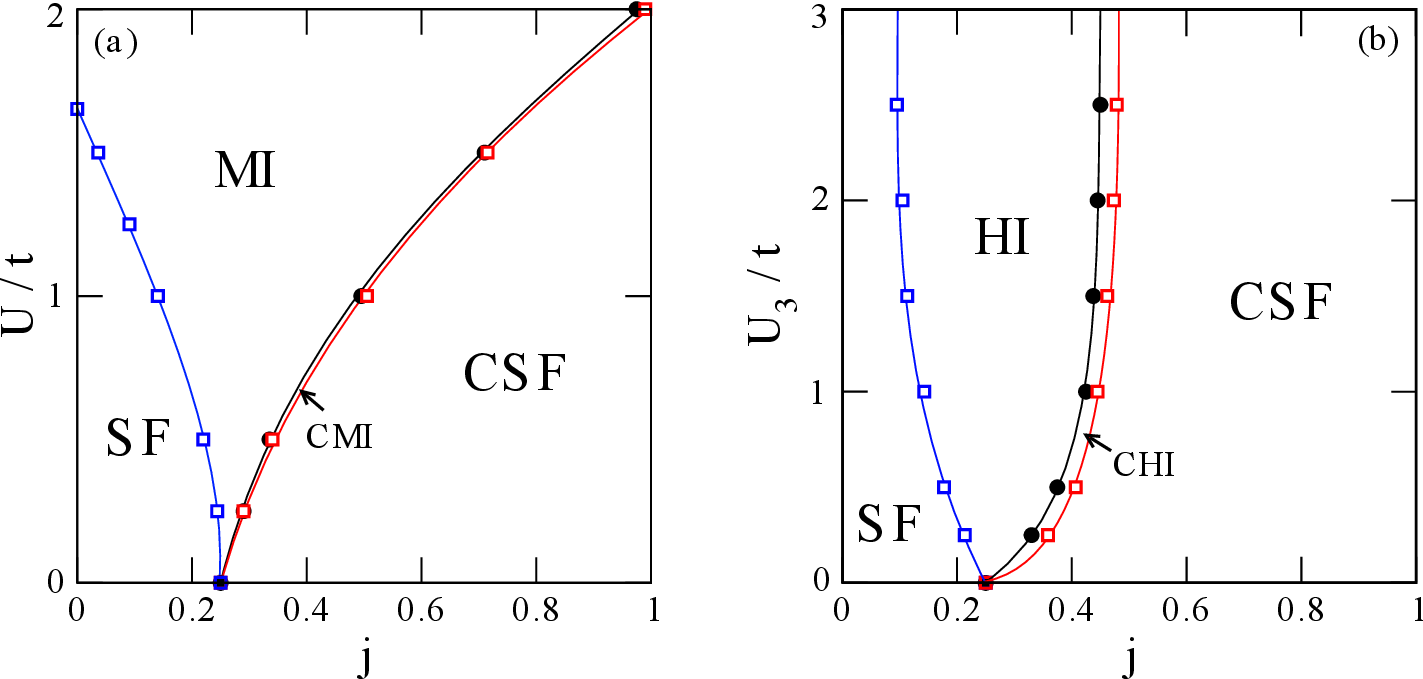}
\caption{Phase diagram for unconstrained bosons as a function of the frustration parameter $j$ and (a.) the on-site interaction $U$~(with $U_3=0$) 
and (b.) the three-body repulsion $U_3$~(and $U=0$). 
In the figures, $\bigcirc$ indicate the boundary of the chiral phases characterized by long-range ordered chirality-chirality correlations $\left<\kappa_i\kappa_j\right>$;  
$\square$ indicate the boundary of the SF-phases indicated by the critical Luttinger parameter $K=2$, which for the SF-MI (SF-HI)-BKT-transition coincide well with the results from energy level crossings denoted by $\times$ (cf. supplementary material B). Note that narrow CMI and CHI phases may occur as well (cf. supplementary material C). The dotted line ($+$-symbols) marks the Lifshitz-line separating commensurate and incommensurate behaviour of the system. Here momentum-distribution splits into two inequivalent maxima.
}
\label{fig:2}
\end{figure}

To study the effect of interactions we combine numerical calculations based 
on the density matrix renormalization group~(DMRG) method~\cite{White} (with up to $N=300$ 
sites keeping per block on average $400$ states for gapped phases and $600$ states 
for gapless ones), and bosonization 
techniques to unveil the low-energy behavior of model~\eqref{eq:H-BH}. 
For $j<1/4$, we employ standard bosonization transformations~\cite{Giamarchi}, 
with an additional oscillating factor $b_i\to (-1)^ie^{i\sqrt{\pi}\theta(x)}$,
to obtain the low-energy effective theory, which is given by the sine-Gordon model
\begin{equation}
\label{sine-Gordon}
{\mathcal H}\!=\! \frac{v_s}{2}\left[ \frac{(\partial_x \phi)^2}{K}+
K(\partial_x \theta)^2  \right]\!-\!{\mathcal M} 
\cos[2\pi \bar n x-\sqrt{4\pi} \phi ],
\end{equation} 
where $\theta$ and $\partial_x\phi$ describe phase and density fluctuations of bosons respectively, $[\theta(x),\partial_y\phi]=i\delta(x-y)$, $v_s$ is the sound velocity and  $K$ the Luttinger parameter. 
In the weak-coupling, $Um\ll 1$, hydrodynamic relations 
are expected to hold: $v_s(j)\sim \sqrt{\bar n U/m\pi^2}=v_s(0)\sqrt{1-4j}$ and $K(j)\sim \sqrt{\bar n\pi^2/ Um}=K(0) \sqrt{1-4j}$, clearly showing that $j$ enhances correlations.
At $j=1/4$, $m$ diverges and the system enters a Mott-insulator~(MI) 
even for vanishingly small $U$~ (Fig.~\ref{fig:2}(a)). The SF-MI transition 
takes place however in the strong-coupling regime in which $v_s$ and $K$ must 
be determined numerically. We obtain $K$ from the single-particle correlations
$G_{ij}  =\langle b_i b^{\dagger}_j\rangle$ which in the SF decay as 
$\sim (-1)^{i-j}{|i-j|^{-1/2K}}$. 
The value $K=2$ marks the boundary between SF~($U<U_c$, $K>2$) and 
MI~($U>U_c$, $K<2$, and ${\mathcal M}>0 $). The MI phase is 
characterized by a hidden parity order~\cite{Berg2008}, 
${\cal O}^2_P=\lim_{|i-j|\rightarrow\infty}\langle 
(-1)^{\sum_{i<l<j}\delta n_l}\rangle\sim\langle \cos \sqrt{\pi} 
\phi \rangle^2$, which has been recently measured in site-resolved 
experiments~\cite{Endres2011}.

The $j>1/4$ case is best understood from bosonization in the $j\gg 1$ regime. We may then 
introduce two pairs of bosonic fields ($\theta_{1},\phi_{1}$) 
and ($\theta_{2},\phi_{2}$), describing, respectively, the subchains 
of even and odd sites. The effective model is governed by 
the Hamiltonian density
\begin{eqnarray}
\label{effectivetwocomponent}
{\mathcal H}&=&\sum_{\alpha=\pm}\frac{ v_{\alpha}}{2}\left[ \frac{(\partial_x \phi_{\alpha})^2}{ K_{\alpha}}+ K_{\alpha}(\partial_x \theta_{\alpha})^2  \right]  \\
&+& \lambda\partial_x \theta_+\sin \sqrt{2\pi}\theta_--  2{\mathcal M} \cos\sqrt{2\pi} \phi_{+}  \cos\sqrt{2\pi} \phi_{-}\nonumber
\end{eqnarray}
where $\theta_\pm=(\theta_1\pm \theta_2)/\sqrt{2}$, $\phi_\pm=(\phi_1\pm \phi_2)/\sqrt{2}$, 
$v_{\pm}$, $K_{\pm}$, and  ${\mathcal M}$ are phenomenological parameters (in the regimes displayed on Figure 2 (a)), $\lambda\sim j^{-1}$. 
Note that the chirality is given by $\kappa_i\to \sin \sqrt{2\pi}\theta_-(x)$.
In weak-coupling, $Um'\ll1$, with $m'=(\partial^2 \epsilon(k)/\partial k^2)_{k_0}^{-1}=4j/t(16j^2-1)$, $v_{\pm}\sim \sqrt{\bar n U/m'\pi^2}$ and $K_{\pm}\sim \sqrt{\bar n\pi^2/ Um'}$. In this case only the term $\partial_x \theta_+\sin \sqrt{2\pi}\theta_-$ is relevant, resulting in 
$\langle \sin \sqrt{2\pi}\theta_-\rangle\neq 0$~\cite{Nersesyan}. 
Hence, a small $U$ is expected to favor a CSF for $j>1/4$, as our numerical results confirm~(Fig.~\ref{fig:2}(a)). 
The CSF phase is characterized by $G_{ij}\sim (-1)^{i-j}e^{-i\kappa(i-j)}{|i-j|^{-1/4 K_+}}$, where $\kappa\sim \langle \kappa_i\rangle$.

Moreover, depending on the values of $K_{\pm}$ bosonization opens the possibility of two consecutive phase transitions 
with increasing $U$ starting from the CSF phase~(App.~\ref{App:A}), which we have confirmed with our DMRG calculations~(Fig.~\ref{fig:2}(a)) detailed in App.~\ref{App:B}. 
First a KT transition occurs from CSF to chiral-Mott~(CMI), a narrow Mott phase with finite chirality. Then 
an Ising transition is produced from CMI to non-chiral MI.
At both KT transition lines in Fig.~\ref{fig:2}(a) (SF-MI and CSF-CMI), up to a logarithmic prefactor, 
$G_{ij}\sim (-1)^{i-j}e^{-i \kappa(i-j)}{|i-j|^{-1/4}}$, where in CSF $ \kappa\neq 0$.

\section{Constrained bosons} 
\label{sec:Constrained}
As mentioned above, sufficiently large three-body losses may result in a three-body constraint $(b_i^\dag)^3=0$~($U_3=\infty$)~\cite{Daley2009}. 
In that case, Model~\eqref{eq:H-BH} may be mapped to a large extent onto a frustrated spin-$1$ chain model~\cite{commentSpin1}, which, presents 
the possibility of a gapped Haldane phase, characterized by a non-local string order. Hence, interestingly, constrained bosons in a zig-zag lattice 
may be expected to allow for the observation of the HI phase in the absence of polar interactions.

Indeed, a model with $U=0$ and finite $U_3$ shows that at the Lifshitz point, $j=1/4$, a HI phase is stabilized for arbitrarily weak $U_3$~(Fig.~\ref{fig:2}(b)).
The effective theory describing the HI is again the sine-Gordon model~(\ref{sine-Gordon}) with $K<2$. However, now ${\cal M}<0$, which selects a hidden string order ${\cal O}^2_S\equiv\lim_{|i-j|\rightarrow\infty}\langle \delta n_i \exp [i\pi\sum_{i<l<j}\delta n_l]\delta n_j\rangle\sim \langle \sin \sqrt{\pi}\theta\rangle^2$~\cite{Berg2008}. Resembling the case of Fig.~\ref{fig:2}(a), SF, HI, chiral-HI~(CHI) and CSF phases occur~(Fig.~\ref{fig:2}(b)). 
These phases are expected for $U_3=\infty$ from known results in frustrated spin-$1$ chains~\cite{Kolezhuk,Lecheminant,Hikihara2000,Hikihara2002}. 
Our DMRG simulations suggest that all these phases meet at $j=1/4$ for $U_3\to 0$. 

\begin{figure}[t]
\includegraphics[width=8.0cm]{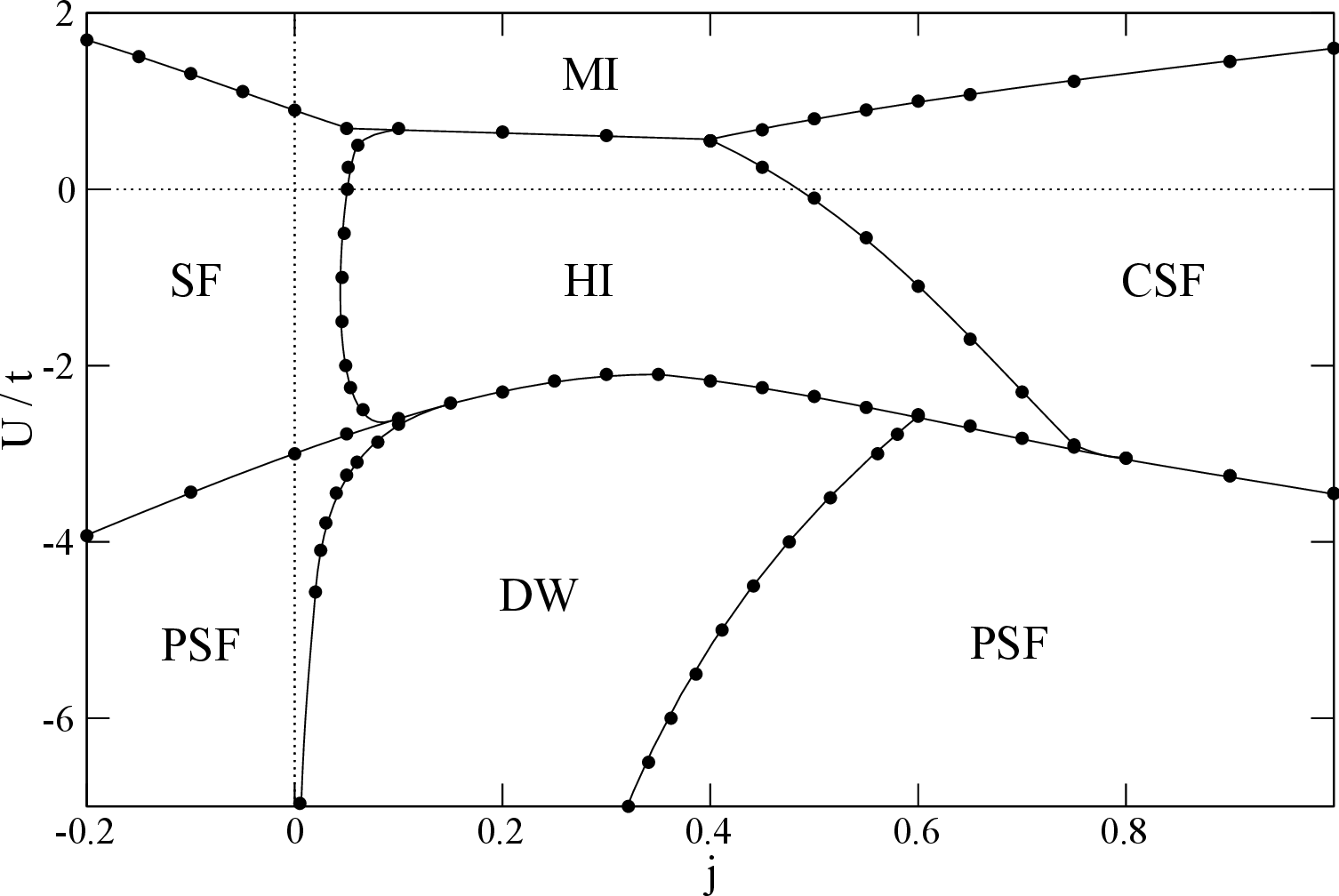}
\caption{Phase diagram for constrained bosons as a function of the on-site interaction $U$ and the frustration parameter $j$.  
Narrow CMI and CHI phases 
may occur along the phase transition lines from CSF to MI and CSF to HI, respectively, but their extension would be negligible in the figure. 
For the precise location of the PSF-DW and HI-MI transition we have additionally analyzed the energy-level-crossings with, respectively, periodic and twisted-boundary conditions. The KT transitions from SF to MI and HI have been determined by the extraction of the Luttinger parameter $K$~\cite{commentLuttinger}.}
\vspace*{-0.3cm}
\label{fig:3}
\end{figure}

Figure~\ref{fig:3} shows the phase diagram for constrained bosons~($U_3=\infty$). 
Starting from the HI phase, increasing $U>0$ can induce a Gaussian HI-MI phase transition, characterized by a vanishing ${\cal M}=0$ in \eqref{sine-Gordon}, 
resembling the phase transition between Haldane and large-D phases induced by single-ion anisotropy in spin-1 chains~\cite{Schulz}. The SF phase is separated from the MI and HI by KT transitions, whereas at the CSF boundary with the MI~(HI) a CMI~(CHI) occurs as mentioned above~(these very narrow regions are not resolved in Fig.~\ref{fig:3}). 

Interestingly, constrained bosons allow as well for the exploration of attractive two-body interactions, $U<0$, without collapse. The $U<0$ phases are also depicted in Fig.~\ref{fig:3}. For sufficiently large $|U|$, bosons tend to cluster in pairs and, as already discussed in Ref.~\cite{Daley2009}, for $j=0$ an Ising transition between 
a SF and a pair superfluid~(PSF) occurs\cite{commentIsing}, analogous to the XY1 to XY2 phase transition in spin-$1$ chains induced by single-ion anisotropy~\cite{Schulz}
~(this transition has been recently studied for 2D lattices as well~\cite{Bonnes2011, Chen2011}). The PSF phase is characterized by an exponentially decaying $G_{ij}$ but algebraically
decaying pair-correlation function $G_{ij}^{(2)}=\langle (b_i^\dag)^2(b_j)^2\rangle$. 
Indeed a PSF occurs for sufficiently large $|U|$, also for $j>1/4$ which is characterized in bosonization in Eq.~\eqref{effectivetwocomponent} by a gapped antisymmetric sector, with pinned $\phi_-$, and a gapless symmetric sector~\cite{Vekua}. 
Though one may anticipate an Ising phase transition between the CSF~(with broken discrete parity symmetry)  and the 
PSF~(with restored symmetry), the behavior of ${\cal O}_P^2$ and $\kappa$~(not shown) hints to a weakly first-order nature.

Small $U<0$ disfavors singly-occupied sites and thus enhances ${\cal O}_S^2$ and 
the bulk excitation gap of the HI phase~(see Figs.~\ref{fig:4} and \ref{fig:5}). However, since 
large $U<0$ removes singly occupied sites completely, just like strong nearest neighbour repulsion, it is expected that the HI phase eventually will transform for growing $|U|$ into a gapped density-wave~(DW) phase via Ising phase transition~\cite{DallaTorre2006}, and string order will evolve into DW order (Fig.~\ref{fig:4} shows how ${\cal O}_s$ merges with 
${\cal O}_{DW}\equiv \lim_{j\rightarrow\infty}(-1)^j \langle n_in_{i+j}\rangle$ for $U/t<-3$). The DW phase is characterized by an exponential decay of both $G_{ij}$ and $G_{ij}^{(2)}$ though a finite ${\cal O}_{DW}$.
 Our DMRG results confirm this scenario (see Fig.~\ref{fig:4}), showing that a DW phase is located between the above mentioned PSF regions~(Fig.~\ref{fig:3}).  

Interestingly the DW phase remains in between both PSF regions all the way into $U\rightarrow-\infty$. In that regime, we may project out 
singly-occupied sites, and introduce a pseudo-spin-$1/2$, identifying $|0 \rangle\to|\!\!\downarrow \rangle ,  |2\rangle\to|\!\!\uparrow \rangle $, and defining the spin operators $\tau_i^-\to (-1)^i b_i^2/\sqrt{2}$, $2\tau_i^z\to b_i^{\dagger}b_i-1$. The 
effective model to leading order in $1/|U|$ is a spin-$1/2$ chain:
\begin{equation}
\label{effectivespinhalf}
H_{\frac{1}{2}}\!=\!J \!\sum_i \!\!\left[{\bs \tau}_i{\bs \tau}_{i+1}\!+\!  j^2(\tau_i^z 
\tau_{i+2}^z\!-\!\tau_i^x \tau_{i+2}^x\!-\!\tau_i^y \tau_{i+2}^y)
\right],
\end{equation}
where $J={t^2}/{|U|}$. For $j=0$, this is a $SU(2)$ symmetric chain, whereas the $j^2J$ terms break the symmetry down to $U(1)$, moving 
the effective theory obtained after bosonization of $H_{\frac{1}{2}}$ towards the irrelevant direction~(in the renormalization group sense). As a result of this, a gapless XY phase of the spin-1/2 chain is expected, i.e. a PSF phase. Higher order terms in $1/|U|$~(not shown explicitely) break, even for $j=0$, the $SU(2)$ symmetry to $U(1)$ in the irrelevant direction. However, interestingly, the ring exchange along the elementary triangle of the zig-zag chain, with amplitude $jt^3 /U^2$, forces the effective theory towards the relevant direction, leading to a gapped N\'eel phase of the spin-1/2 chain, i.e. the DW phase. The competition between exchange along the lattice bonds and ring-exchange leads hence to two consecutive KT phase transitions induced by $j$, for $j\ll 1$ first from PSF to DW, followed by DW back to PSF. The width of the DW phase is $\sim t/|U|$, and it extends all the way into the $U\to -\infty$ limit.

Finally, our DMRG simulations show a narrow region where a direct, apparently first-order, HI-PSF transition occurs (see for details App.~\ref{App:B}), characterized 
by  discontinuous jumps of ${\cal O}_{S,P}^2$~(Fig.~\ref{fig:5}). This first-order nature is explained because on one hand 
increasing $|U|$ within the HI phase increases ${\cal O}_S^2$ due to the suppression of singly-occupied sites, and on the other 
hand, for $0.6 \lesssim j \lesssim 0.75$~(Fig.~\ref{fig:3}), a growing $|U|$ 
destroys the insulating state in favour of a PSF phase, where string order cannot exist. 
On the contrary, ${\cal O}_S^2$ diminishes for decreasing $|U|$ when approaching the HI-CSF boundary~(Fig.~\ref{fig:5}).  

\begin{figure}[t]
\includegraphics[width=8.0cm]{bh3_cut_j0.3.eps}
\caption{Order parameters ${\cal O}^2_P$~($\square$), ${\cal O}^2_S$~($\bigtriangleup$), ${\cal O}_{DW}$~($\bigtriangledown$), and
energy gap~($\times$) as function of $U$ for constrained bosons on the $j=0.3$ line ($N=160$).
The parity, as defined in the Mott state, must get an additional oscillating factor in the DW phase ${\cal O}^2_P \to (-1)^{i-j}{\cal O}^2_P$.}
\vspace*{-0.3cm}
\label{fig:4}
\end{figure}

\begin{figure}[t] 
\includegraphics[width=8.0cm]{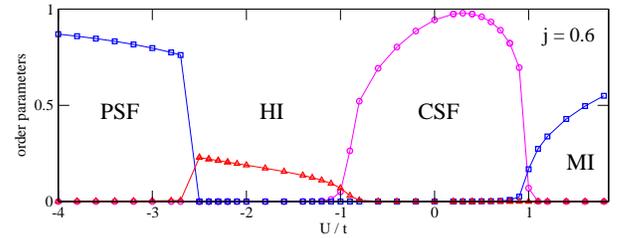}
\caption{Order parameters ${\cal O}^2_P$~($\square$), ${\cal O}^2_S$~($\bigtriangleup$), and $\kappa^2$~($\bigcirc$)
as function of $U$ for constrained bosons on the $j=0.6$ line ($N=160$). Both ${\cal O}_{P,S}^2$   
show jumps at the HI to PSF transition at $U\simeq -2.5$. In the PSF, similar to the DW phase, ${\cal O}_P^2$ is defined with an 
additional oscillating factor.}
\vspace*{-0.3cm}
\label{fig:5}
\end{figure}


\section{Conclusions}
\label{sec:conclusions}
 In conclusion, the interplay between geometrical frustration and interactions leads to rich physics for ultra-cold bosons in zig-zag optical lattices. Unconstrained bosons may present chiral superfluidity, and Mott insulator for vanishingly small interactions. 
Constrained bosons may allow for the observation of Haldane-insulator without the necessity of polar interactions, as well 
as pair-superfluid and density wave phases at attractive interactions. All the predicted phases may be detected 
using state of the art techniques. The SF and CSF phases may be distinguished by means 
of time-of-flight~(TOF) techniques, in a similar way as recently done for condensates in triangular optical lattices~\cite{Struck2011}.
The DW and PSF phases are characterized by double or zero occupancy, which could be detected using parity measurements as those 
introduced in Refs~\cite{Bakr2010,Sherson2010}, and could be discerned from each other by the absence/presence of interference fringes in 
TOF~\cite{Greiner2002}. Finally, the string-order of the HI phase may be studied using similar site-resolved measurements as those recently reported for the measurement of non-local parity order in Mott insulators~\cite{Endres2011}. 

\acknowledgements
This work has been supported by the cluster of excellence QUEST (Center for Quantum Engineering and Space-Time Research). T.V. acknowledges SCOPES Grant IZ73Z0-128058.

\appendix 

\section{ Bosonization analysis of the MI-CMI-CSF transition }
\label{App:A}

\renewcommand\thefigure{A\arabic{figure}}
\setcounter{figure}{0}

In this appendix we provide additional details on our bosonization analysis.
We take the limit $j\gg 1$, that allows to consider sub-chains formed by the even and odd sites, 
and introduce the symmetric and antisymmetric fields $\theta_\pm$ and $\phi_\pm$. The interaction between 
these fields is given by the last two terms of Eq. (3) of the main text:
\begin{equation}
\label{interaction}
{\mathcal H}_{int}
\!=\! \lambda\partial_x \theta_+\sin\! \sqrt{2\pi}\theta_-\!-\!  2{\mathcal M} \cos\!\sqrt{2\pi} \phi_{+}  \cos\!\sqrt{2\pi} \phi_{-},
\end{equation}
where the first term supports chirality and the second one favors a MI phase. 

Starting from $U\gg t'$, deep in the Mott phase of each sub-chain, then $\langle\cos\sqrt{2\pi} \phi_{\pm}\rangle \neq 0$, and thus the 
$\phi_\pm$ fields are pinned in the Mott phase. 
In this case, to second order in $\lambda$, one can integrate out in the partition function 
$\partial_x \theta_+$ from the first term of Eq.\eqref{interaction}, obtaining the following contribution in the antisymmetric sector,
\begin{equation}
 -\frac{\lambda^2 }{2}\int {\mathrm d}^2 x{\mathrm d}^2 y \langle \partial_x \theta_+\partial_y \theta_+ \rangle \sin \sqrt{2\pi}\theta_-(x)\sin \sqrt{2\pi}\theta_-(y)\nonumber
\end{equation}
where the average is performed in the ground state of the MI phase, where $\langle \partial_x \theta_+\partial_y \theta_+ \rangle$ is short ranged. 
Hence the leading contribution in the antisymmetric sector, after carrying out operator product expansion, is 
a term $\sim \lambda^2 \cos {\sqrt{8\pi }\theta_-}$. Note that a $\sim- \lambda^2 (\partial_x \theta_-)^2$ contribution decreasing the value of $K_-$
is obtained as well in the antisymmetric sector.
The competition between $\cos {\sqrt{8\pi }\theta_-}$ and $\cos\sqrt{2\pi} \phi_-$~(obtained using mean-field decoupling of the last term of Eq.\eqref{interaction} 
in the MI phase) is resolved with an Ising phase transition in the antisymmetric sector with increasing $\lambda/U$, 
leading to the pinning of $\theta_-$ in the new ground state $\langle \sqrt{8\pi} \theta_-\rangle=\pi $, so that  $\langle \sin \sqrt{2\pi}\theta_- \rangle \neq 0$, 
driving the symmetric sector into a state with finite topological current, $\langle \partial_x \theta_+ \rangle \neq 0$. 

The simplest scenario to establish the Ising phase transition is for $K_-=2$. 
In that case, performing a mean-field decoupling of the second term in Eq.~\eqref{interaction}, 
the antisymmetric sector is governed by the Hamiltonian density,
$$
{\mathcal H_-}\!\!=\!\! \frac{v_-}{2}\!\!\left[ {(\partial_x \tilde \phi_-)^2}+
(\partial_x \tilde \theta_-)^2  \right]\!-\!{\tilde M} 
\cos\!\sqrt{4\pi} \tilde \phi_-  +\tilde \lambda \cos\!\sqrt{4\pi} \tilde \theta_-,\nonumber 
$$
where $\tilde \theta_-=\sqrt{K_-}\theta_-$, $\tilde \phi_-=\phi_-/ \sqrt{K_-}$, $\tilde M= 2{\mathcal M} \langle \cos\sqrt{2\pi} \phi_{+} \rangle$, and $\tilde \lambda \sim \lambda^2$. The antisymmetric sector can be hence described by two free massive Majorana fermions, 
with masses $\tilde \lambda \pm \tilde M$. At the Ising phase transition, $\tilde \lambda =\pm \tilde M$, and the mass of one of the Majorana fermions vanishes~\cite{Gogolin1998}. 

However, the Mottness of the ground state after the chirality gets long-range ordered, $\langle \sqrt{2 \pi} \theta_-\rangle\neq0$, does not necessarily 
disappear immediately, due to the possibility of a relevant contribution in the symmetric sector, $\cos\sqrt{8\pi }\phi_+$, for $K_+<1$ which stems after integrating out the $\cos{\sqrt{2\pi} \phi_-}$ in the last term of Eq.~\eqref{interaction} in the state with pinned $\theta_-$. Note that in the 
CMI state $\langle \theta_-\rangle\neq 0$, $\langle \partial_x \theta_+ \rangle \neq 0$, and also $\langle\phi_+\rangle\neq 0$. 
Further decreasing $U$, at $K_+=1$, the CMI phase~($K_+<1$) disappears at a KT phase transition in favor of the CSF phase~($K_+>1$).

Thus, our bosonization analysis, for $j\gg 1$, suggests the possibility of two consecutive phase transitions with increasing $U$ starting from the CSF phase, first a KT transition from CSF to CMI, 
followed by an Ising transition from CMI to MI. Note that if an intermediate CMI phase 
between CSF and MI is absent, the direct transition between CSF and MI cannot be of Ising nature, 
since CSF is a gapless phase and MI is gapped. In the following section, we show provide a numerical 
proof of the existence of the CMI phase, showing that starting from the MI phase, the system experiences 
an Ising transition involving the growth of chirality. 

\section{ Numerical analysis}
\label{App:B}
Here  we provide details on our numerical calculations, and in particular on how phase boundaries were determined and how error bars were estimated. 

We investigate the phase diagrams by means of numerical simulations based on exact diagonalization and density-matrix renormalization group (DMRG) with up to $300$ lattice sites. Typically for DMRG-simulations we keep about $\chi\approx 500$ matrix-states. The results have also been confirmed by infinite-system size algorithm~(iDMRG)~\cite{Schollwock2005} with up to $\chi\approx 300$ states. For our simulations of the unconstrained Bose-Hubbard-model we kept $n_{max}=4$ bosons per site for $U\geq 1$ and $n_{max}=6$ for $U=0.5$, which has been shown to be sufficient by comparing to simulations with higher $n_{max}$. For open-boundary-conditions special care may be needed to take care of degenerate edge states in the Haldane-phase, i.e. by polarizing edges~\cite{DallaTorre2006}.

\begin{figure}[tb]
\includegraphics[width=\columnwidth]{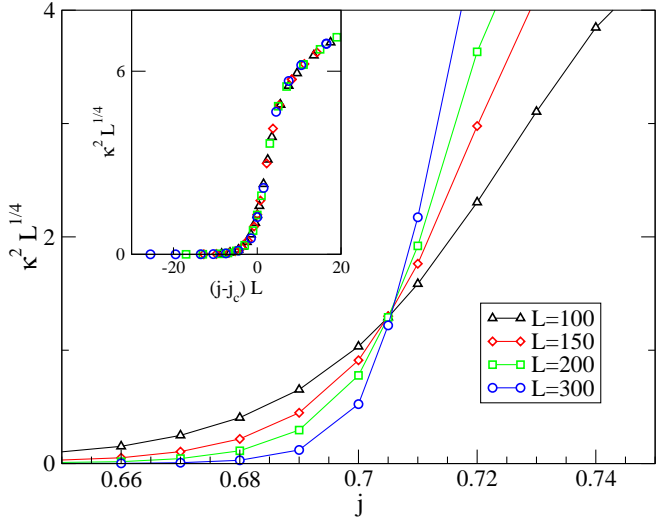}
\caption{ (Color online)  Chirality scaling $\kappa^2 \cdot L^{1/4}$ close to MI-CMI-transition for $U=3$. The inset shows the collapse of all finite system size data to a single curve. The results indicate a MI-CMI Ising transition at $j=0.705\pm 0.001$.}
\label{fig:C1}
\end{figure}

First we discuss phase transitions involving chiral order parameter $\langle \kappa_i \rangle$, where 
$\kappa_i$ is defined in Sec.~\ref{sec:Unconstrained}.
Since chiral order is spontaneous, we study numerically the chirality-chirality correlation function $\langle \kappa_i \kappa_{i+n}\rangle$~\cite{Hikihara2000}, which at large distances saturates to $\kappa^2$ and we extract it by averaging over long distances. We study how the chirality vanishes with changing $j$ 
(for the MI-CMI and HI-CHI transitions; for the CSF-PSF transition we study instead the behavior of the parity order). The scaling of chirality $\kappa^2 \cdot L^{1/4}$ close to MI-CMI and HI-CHI transitions unambiguously confirms that the corresponding phase transition is of Ising type, showing the correct scaling behavior. The critical value of $j$ for the corresponding Ising transition is located by extracting the intersection of curves for different system sizes $L$ (the width of the intersecting point provides the uncertainty of the procedure). Fig.~(\ref{fig:C1}) illustrates our numerical results in the vicinity of the MI-CMI transition. The collapse of the data for different system sizes on a single curve~(inset) confirms the Ising nature of the underlying phase transition. Similarly we confirm numerically the Ising nature of the HI-CHI transition~(not shown).

\begin{figure}[tb]
\includegraphics[width=\columnwidth]{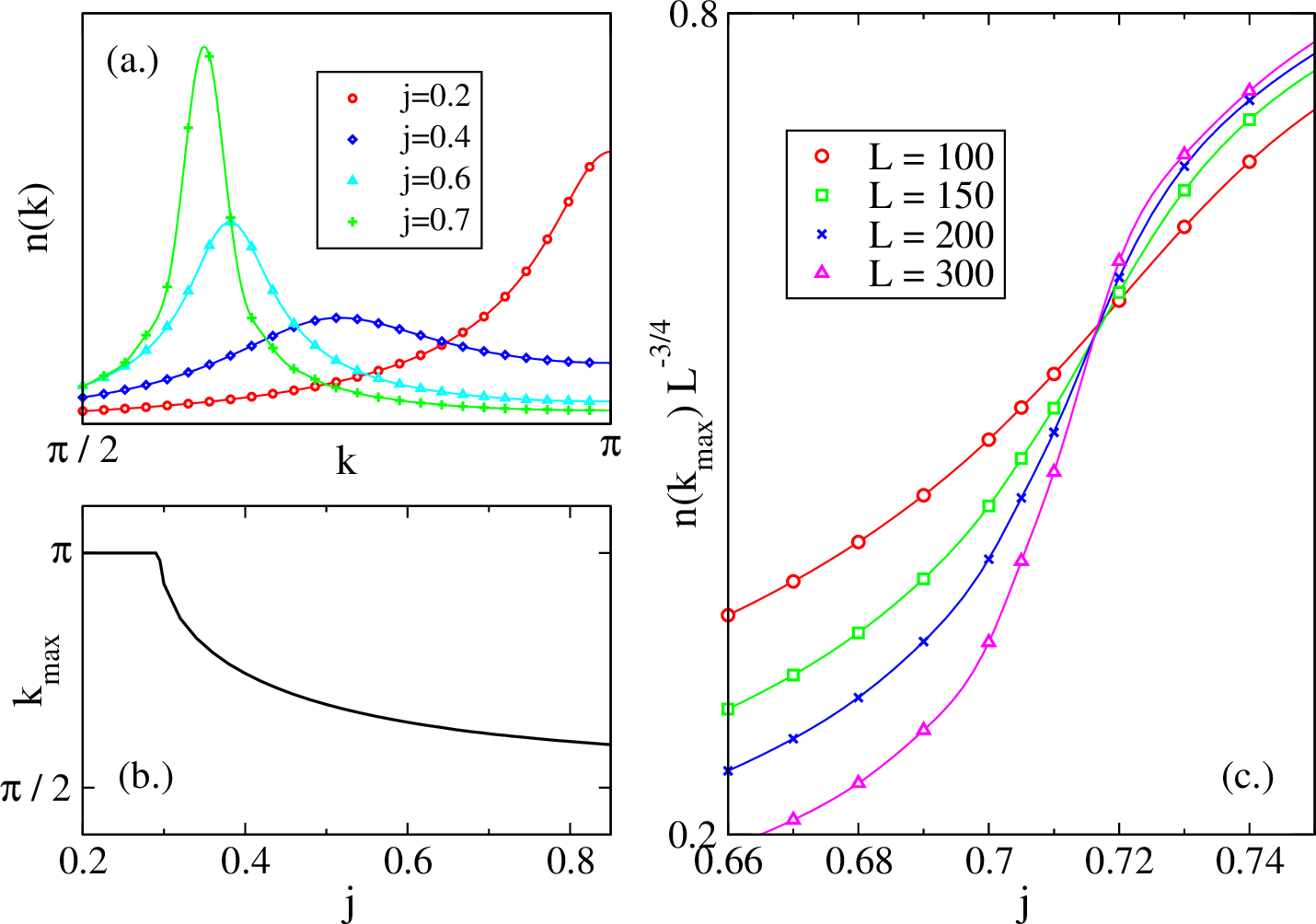}
\caption{ (Color online) (a) Momentum distribution $n(k)$ for unconstrained bosons, for $U=3$ and $L=100$ (the position of the maximum $k_{max}$ is practically independent of $L$).(b) Value of $k_{max}$ as a function of $j$. At $j\approx 0.3$ the quasi-momentum peak departs from $\pi$ and $n(k)$ acquires two inequivalent maxima at $\pm k_{max}$. (c) Scaling of $n(k_{max}) L^{-3/4}$ for different $L$. The crossing of the curves indicates the CMI-CSF transition at $j= 0.717\pm0.001$. Extracting the Luttinger liquid parameter as indicated in the text gives a similar estimate of $j= 0.72\pm 0.01$ though with an order of magnitude larger error bar.}
\label{fig:C2}
\end{figure}

The standard numerical procedure to locate the KT transition from superfluid to gapped phase is based on the Luttinger liquid parameter, which we extract from the single-particle correlation function $G_{ij}$. When the single particle correlations show incommensurate oscillations, we fit it to $G_{ij}\propto G(r)^{1/2K} \cos(\omega x +\phi)$ with $G(r)$ including conformal corrections\cite{Cazalilla2004}. To get a lower bound for the transition point one can apply a power law fit to shorter distances after dividing out incommensurate oscillations. The KT CMI-CSF transition (as well as the CHI-CSF transition) can be located in this way providing a strong hint on the existence of a finite CMI~(CHI) region between the MI~(HI) and CSF-phases. However, a much more accurate estimate of the KT transition point is provided by the analysis of the quasi-momentum distribution $n(k)$~\cite{Dhar2011}. Since $G_{ij}\sim e^{-i Q (i-j)}{|i-j|^{-\alpha}}$ with $\alpha=1/4$ at the KT transition (up to logarithmic correction), $n(k)$ has a maximum at $Q$ and its height depends on the system size as $n(Q) \approx \frac{1}{L} \sum_{i,j} e^{iQ (i-j)} G_{ij} \sim L^{1-\alpha}$. This behavior is illustrated in Fig.~\ref{fig:C2}, which shows a clear intersection of the $n(k_{max})\cdot L^{\alpha-1}$ curves for different system sizes at a single point~(the width of this point provides the error bar for the position of the KT transition). Hence for the case depicted in 
Figs.~\ref{fig:C1} and~\ref{fig:C2}, a narrow but clearly determined CMI phase may be found between $j=0.705\pm 0.001$ and $j=0.717\pm 0.001$.

In addition to chirality we have studied other order parameters, including parity, string, and DW orders~(all defined in the main text). Applying finite-size-scaling analysis~\cite{Ueda2008} allows or an accurate 
location of the corresponding phase transition lines depicted in Fig.~3 of the main text.\\

\begin{figure}[b] 
\includegraphics[width=0.45\columnwidth]{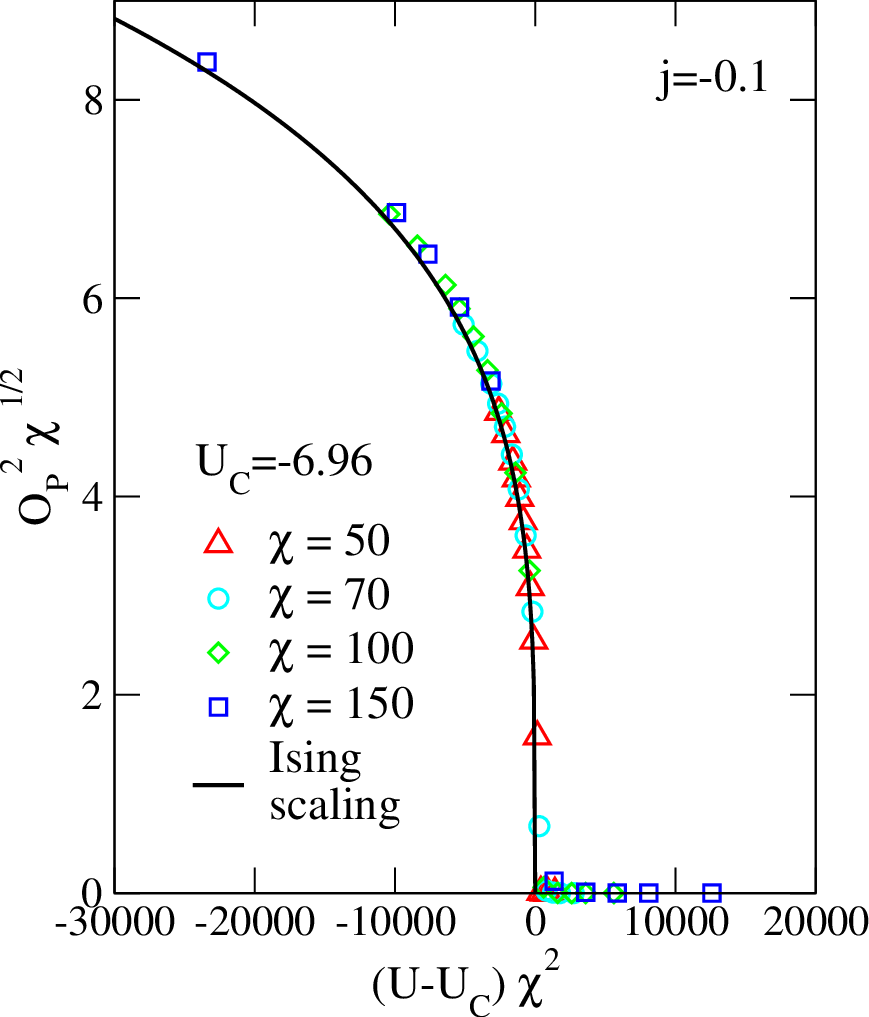}
\includegraphics[width=0.42\columnwidth]{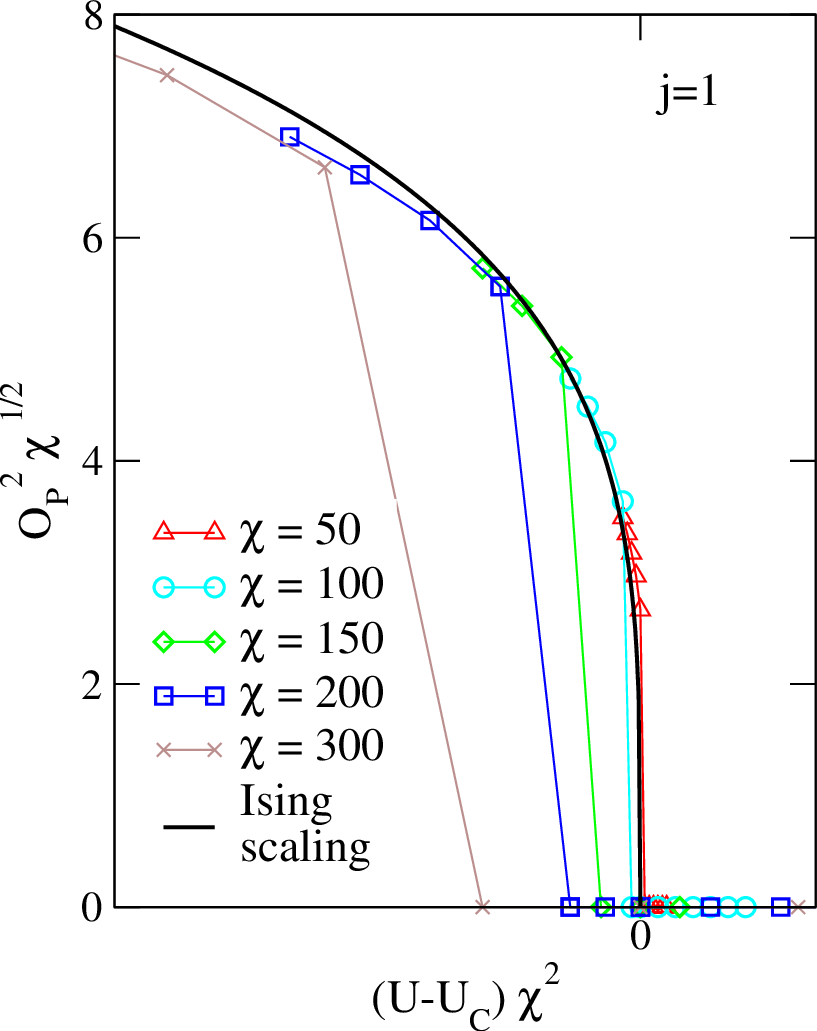}
\caption{ (Color online) (left) Scaling with the matrix dimension $\chi$ of the parity order parameter for the PSF-SF 
transition. Note that the scaling accurately follows the expected one for an Ising transition. (right) 
Same results for the PSF-CSF transition show clear deviations from the expected Ising scaling, exhibiting an abrupt jump in the parity order.}
\label{fig:D1}
\end{figure}

\begin{figure}[tb]
\includegraphics[width=\columnwidth]{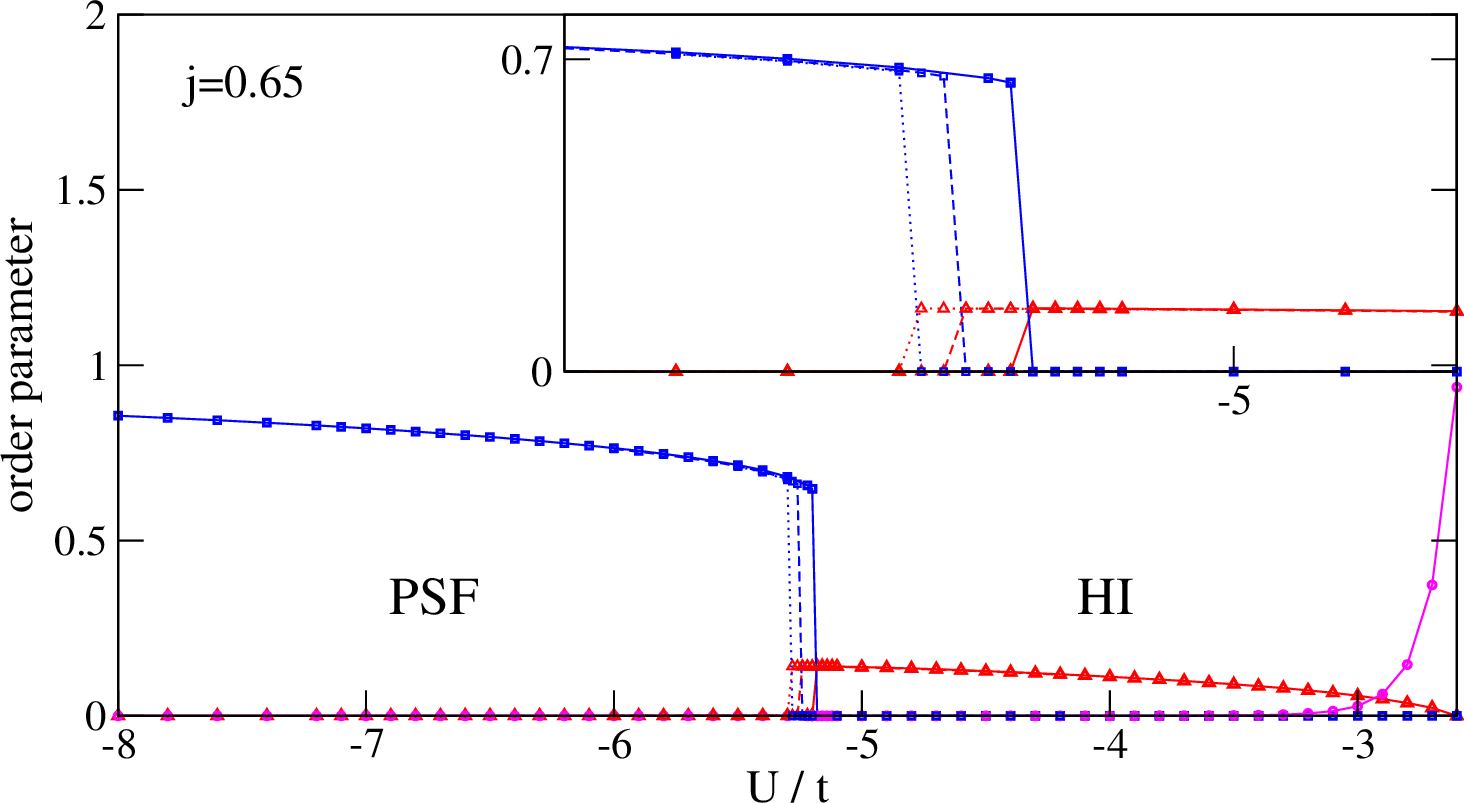}
\caption{ (Color online) Parity-order ${\cal O}^2_P$~($\square$), string-order ${\cal O}^2_S$~($\bigtriangleup$) and chirality $\kappa^2$~($\bigcirc$) for constrained bosons ($j=0.65$) close to the HI - PSF transition of iDMRG calculations for different matrix dimensions: $\chi=100$ (straight line), $\chi=200$ (dashed line), $\chi=300$ (dotted line). The transition point shifts slightly with increasing $\chi$. The inset shows a zoom in the HI-PSF transition region.}
\label{fig:B2}
\end{figure}

On Fig.~\ref{fig:D1} we depict the scaling of the parity order in the vicinity of PSF-SF and PSF-CSF 
transitions using the iDMRG algorithm. The correlation length for the quantum Ising model scales like 
$\xi = \chi^2$, where $\chi$ denotes the matrix dimension~\cite{Tagliacozzo2008}. This result is in very good agreement with our results for the PSF-SF transition~(Fig.~\ref{fig:D1}, left). Indeed, using Ising critical exponents the parity order parameter behaves as
\[
{\cal O}^2_P = \xi^{-1/4} f\left((U-U_c) \xi\right).
\]
On the contrary, for the CSF-PSF transition the behavior is fundamentally different, and we instead observe 
a jump in the parity order~(Fig.~\ref{fig:D1}, right), indicating a first-order character of the transition.
We observe as well a similar jump in the parity order across the PSF-HI transition. In addition string order shows an abrupt jump at the same transition as depicted on Fig.~\ref{fig:B2}.

Besides monitoring order parameters we use a finite-size level crossing analysis~(level spectroscopy)~\cite{LevelSpectroscopy} to determine the location of different phase transitions. We use DMRG for longer chains, which compares well with our exact diagonalization results available only up to $16$ sites. The SF-PSF transition can be extrapolated very precisely from level crossing between the one-particle and two-particle excitation gaps. 
We determine in this way the SF-PSF boundary, which compares well with the boundary obtained from monitoring the parity order. The transitions from DW to PSF phases can also be determined very accurately by level spectroscopy. Finite-size extrapolation following $1/N^2$ law confirms the KT nature of those transitions. Finally, for twisted boundary conditions the Gaussian transition line between HI and MI can also be located by ground state level crossing~\cite{Chen03}.


\begin{thebibliography}{10}
\bibitem{commentIsing}  Numerically we have unambiguously confirmed the Ising nature of SF-PSF transition by observing 
the correct scaling of ${\cal O}_P^2$ close to transition point.
\bibitem{commentLuttinger} It is notoriously hard to determine precisely the KT phase transition lines by this method alone. 
The precise location of the phase transition lines, and the determination of the nature of the multi-critical points 
close to $j=0$ are not pursuit in the current work.
\end{thebibliography}
\end{document}